\documentclass[10pt,twocolumn,twoside]{IEEEtran}
\usepackage[algoruled, linesnumbered]{algorithm2e}
\usepackage{cite}
\usepackage{graphicx}
\usepackage{psfrag}
\usepackage{epsfig}
\usepackage{subfigure}
\usepackage{url}
\usepackage{stfloats}
\usepackage{amsmath}
\usepackage{array}
\usepackage{array}
\begin{document}
\title{In Search of Autocorrelation Based Vocal Cord Cues for Speaker Identification}
\author{\authorblockN{Md. Sahidullah\authorrefmark{1} and
Goutam Saha\\
\authorblockA{Department of Electronics and Electrical Communication
Engineering\\Indian Institute of Technology, Kharagpur, India,
Kharagpur-721 302\\Email: \authorrefmark{1}sahidullah@iitkgp.ac.in, gsaha@ece.iitkgp.ernet.in
\\Telephone:
+91-3222-283556/1470, FAX: +91-3222-255303}}
\\
}
\maketitle

\begin{abstract}
In this paper we investigate a technique to find out vocal source based features from the LP residual of speech signal for automatic speaker identification. Autocorrelation with some specific lag is computed for the residual signal to derive these features. Compared to traditional features like MFCC, PLPCC which represent vocal tract information, these features represent complementary vocal cord information. Our experiment in fusing these two sources of information in representing speaker characteristics yield better speaker identification accuracy. We have used Gaussian mixture model (GMM) based speaker modeling and results are shown on two public databases to validate our proposition.
\end{abstract}

\section{Introduction}
\PARstart{S}peaker Identification \cite{camp,Kinnunen,sandi} is the task to determine the identity of the unknown subject by its voice. It requires a robust feature extraction technique followed by a modeling approach. Through feature extraction process the crude speech signals are undergone through several dimensionality reduction operations where the consequent outputs are more compact and robust than the original speech.  Although the speech signal is a non-stationary signal it shows short term stationary in the interval 20-40 ms \cite{tom}.  The vocal tract characteristics are almost static during this period. Most of the feature extraction techniques are based on short term spectral analysis of the speech signal. The central idea is to capture the information related to the formant frequency and their characteristics like magnitude, bandwidth, and slope etc through various spectrum estimation techniques. The shortcoming of these techniques is that it neglects the vocal cord (or vocal fold) characteristics which also carries significant speaker specific information. The residual signal obtained through linear prediction (LP) analysis of speech contains information related to vocal cord. Some approaches are investigated to find and model the speaker specific characteristics from this residual signal. Auto associative neural network (AANN) \cite{mahadeva}, wavelet octave coefficients of residue (WOCOR) \cite{nengheng}, residual phase \cite{ksrmurty}, higher order cumulant \cite{chetouani}  are employed earlier to reduce the equal error rate (EER) for speaker recognition system. In this work we have considered the autocorrelation method in finding the speaker specific traits from it. The contribution of this residual feature is fused with the spectral feature based system. We have conducted experiments on two popular publicly available speaker identification database (POLYCOST and YOHO) using GMM based classifier. It is observed that the performances of the speaker identification systems based on various spectral features are improved in combined mode for different model order of GMM.
The rest of the paper is organized as follows. In section II we briefly review the basic LP analysis followed by the proposed feature extraction technique. The speaker identification experiment with results is shown in section III. Finally, the paper is concluded in section IV.

\section{Feature Extraction From Residual Signal}

\subsection{Linear Prediction Analysis and Residual Signal}\label{lpccomp}
In the LP model, $(n-1)$-th to $(n-p)$-th samples of the speech wave ($n$, $p$ are integers) are used to predict the $n$-th sample. The predicted value of the $n$-th speech sample \cite{atal} is given by

\begin{equation}\label{lp1}
\hat{s}(n)=\sum_{k=1}^{p}a(k)s(n-k)
\end{equation}

where $\{a(k)\}_{k=1}^{p}$ are the predictor coefficients and $s(n)$ is the $n$-th speech sample.The value of $p$ is chosen such that it could effectively capture the real and complex poles of the vocal tract in a frequency range equal to half the sampling frequency.The prediction coefficients (PC) are determined by minimizing the mean square prediction error \cite{camp} and the error is defined as
\begin{equation}\label{lp2}
E(n)=   \frac{1}{N}\sum_{n=0}^{N-1}(s(n)-\hat{s}(n))^{2}
\end{equation}
where summation is taken over all samples i.e., $N$.
The set of coefficients $\{a(k)\}_{k=1}^{p}$ which minimize the mean-squared prediction error are obtained as solutions of the set of linear equation
\begin{equation}\label{lp3}
\sum_{k=1}^{p}\phi(j,k)a(k)=\phi(j,0) , j=1,2,3,\ldots,p
\end{equation}
where
\begin{equation}\label{lp4}
\phi(j,k)=\frac{1}{N}\sum_{n=0}^{N-1}s(n-j)s(n-k)
\end{equation}

The PC, $\{a(k)\}_{k=1}^{p}$ are derived by solving the recursive equation (\ref{lp3}).

Using the $\{a(k)\}_{k=1}^{p}$ as model parameters, equation (\ref{lp5}) represents the fundamental basis of LP representation. It implies that any signal can be defined by a linear predictor and its prediction error.

\begin{equation}\label{lp5}
s(n)=\sum_{k=1}^{p}a(k)s(n-k)+e(n)
\end{equation}

The LP transfer function can be defined as,
\begin{equation}\label{lp6}
H(z)=\frac{G}{1 - \sum_{k=1}^{p}a(k)z^{-k}}=\frac{G}{A(z)}
\end{equation}
where $G$ is the gain scaling factor for the present input and $A(z)$ is the $p$-th order inverse filter.
These LP coefficients itself can be used for speaker recognition as it contains some speaker specific information like vocal tract resonance frequencies, their bandwidths etc. However, some nonlinear transformations are applied to those PC to improve the robustness. Linear prediction cepstral coefficient (LPCC), line spectral pairs frequency (LSF), log-area ratio (LAR) etc are such representations of LPC.

The prediction error i.e., $e(n)$ is called Residual Signal and it contains all the complementary information that are not contained in the PC. Its worth mentioning here that residual signal conveys vocal source cues containing fundamental frequency, pitch period etc.

\subsection{Autocorrelation of Residual Signal}\label{lagauto}
The autocorrelation finds out the similarity of a signal with itself. It has a beautiful relationship with the power spectral density (PSD). In speech processing tasks the autocorrelation function is mostly popular in estimating the pitch of the signal and in LP based speech analysis. In pitch detection algorithm we compute fundamental frequency as the difference between the two consecutive peaks of autocorrelation function. On the other hand in LP analysis we figure out the second order relationship among the speech samples through autocorrelation function which is previously described in Sec. \ref{lpccomp}.
\\
The autocorrelation of a discrete signal $x(n)$ of length $N$ is given by,
\begin{equation}\label{eqauto1}
r(n) = \sum\limits_{k =  - N}^N {x(k)x(k - n)}
\end{equation}
In eqn. (\ref{eqauto1}), we consider the full correlation over the shift from $-N$ to $N$. The correlation also can be computed with some lag in the original signal to check only the short time similarity. If the maximum and minimum lag is bounded by magnitude of $L$ then the correlation can be calculated as in eqn. (\ref{eqauto2}),
\begin{equation}\label{eqauto2}
r_l (n) = \sum\limits_{k =  - L}^L {x(k)x(k - n)}
\end{equation}
Higher lag autocorrelation of speech signal was employed earlier for robust speech recognition engine \cite{Shannon}.   In our proposed method of feature extraction we have computed the autocorrelation function for a specific lag of residual signal. Firstly, the residual signal is normalized between -1 and +1 to make the first order central moment i.e. mean to zero. This process also helps to reduce the unwanted variation in the autocorrelation. Secondly, we compute the autocorrelation and take the upper half only (due to the symmetrical property of the correlation function). The $r_l(0)$ is also included because it is related to the stress (energy) of a particular speech frame's residual. If we consider the lag of [$-L$, $L$] then total number of coefficients becomes $L+1$. The residual feature extracted through this technique is referred to as ACRLAG throughout this paper.

\subsection{Integration of  Complementary Information}\label{fusion}

\begin{figure*}
\centering
\includegraphics[height =12 cm, width=14 cm]{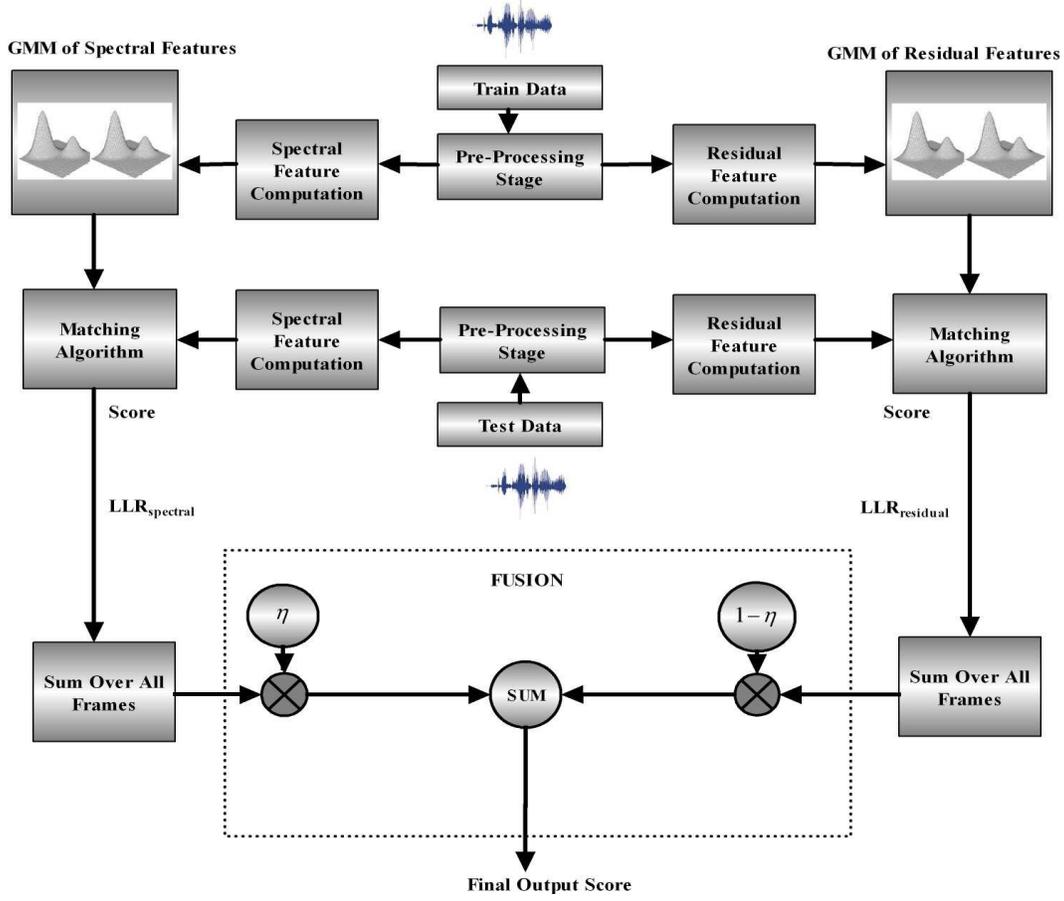}\\
\caption{Block diagram of Fusion Technique: Score level fusion of Vocal tract (short term spectral based feature) and Vocal cord information (Residual).}\label{bdfusion}
\centering
\end{figure*}

In this section we propose to integrate vocal tract and vocal chord parameters identifying speakers. In spite of the two approaches have significant performance difference, the way they represent speech signal is complementary to one another. Hence, it is expected that combining the advantages of both the feature will improve \cite{kittler} the overall performance of speaker identification system. The block diagram of the combined system is shown in Fig. \ref{bdfusion}. Spectral features and Residual features are extracted from the training data in two separate streams.
Consequently, speaker modeling is performed for the respective features independently and model parameters are stored in the model database. At the time of testing same process is adopted
for feature extraction. Log-likelihood of two different features are computed w.r.t. their corresponding models. Finally, the output score is weighted and combined.

We have used score level linear fusion which can be formulated as in equation (\ref{fusioneq}). To get the advantages of both the system and their complementarity the score level linear fusion can be formulated as follows:
\begin{equation}\label{fusioneq}
    LLR_{combined}=\eta LLR_{spectral}+(1-\eta) LLR_{residual}
\end{equation}
where $LLR_{spectral}$ and $LLR_{residual}$ are log-likelihood ratio calculated from the spectral and residual based systems, respectively. The fusion weight is decided by the parameter $\eta$.

\section{Speaker Identification Experiment}
\subsection{Experimental Setup}\label{ee}

\subsubsection{Pre-processing stage}
In this work, pre-processing stage is kept similar throughout different features extraction methods. It is performed using the following steps:
\begin{itemize}
  \item Silence removal and end-point detection are done using energy threshold criterion.
  \item The speech signal is then pre-emphasized with $0.97$ pre-emphasis factor.
  \item The pre-emphasized speech signal is segmented into frames of each $20\mathrm{ms}$ with $50\%$ overlapping ,i.e. total number of samples in each frame is $N = 160$, (sampling frequency $F_s = 8kHz$).
  \item In the last step of pre-processing, each frame is windowed using hamming window given equation
\begin{equation}\label{hamming}
w(n)=0.54-0.46\cos(\frac{2\pi n}{N-1})
\end{equation}
where $N$ is the length of the window.
\end{itemize}

\subsubsection{Classification \& Identification stage}
GMM technique is used to get probabilistic model for the feature vectors of a speaker. The idea of GMM is to use weighted summation of multivariate gaussian functions to represent the probability density of feature vectors and it is given by

\begin{equation}\label{gmm1}
p(\textbf{x})=\sum_{i=1}^{M}p_i b_i(\textbf{x})
\end{equation}
where $\textbf{x}$ is a $d$-dimensional feature vector, $b_i(\textbf{x})$, $i=1,...,M$ are the  component densities and $p_i$, $i = 1,...,M$ are the mixture weights or $\emph{prior}$ of individual gaussian. Each component density  is given by
\begin{equation}\label{gmm2}
b_i(\textbf{x})=\frac{1}{(2\pi)^{\frac{d}{2}}|\mathbf{\Sigma_{i}}|^{\frac{1}{2}}}\exp \bigg\{ -\frac{1}{2}( \textbf{x} - \boldsymbol{\mu_i)}^{t}\mathbf{\Sigma_{i}}^{-1}( \textbf{x} - \boldsymbol{\mu_i})\bigg\}
\end{equation}
with mean vector $\boldsymbol{\mu_i}$ and covariance matrix $\mathbf{\Sigma_{i}}$. The mixture weights must satisfy the constraint that $\sum_{i=1}^{M}p_{i}=1$ and $p_{i}$ $\geq$ $0$.
The GMM is parameterized by the mean, covariance and mixture weights from all component densities and is denoted by
\begin{equation}\label{gmm3}
\lambda=\{ p_{i},\boldsymbol{\mu_i},\mathbf{\Sigma_{i}} \}_{i=1}^{M}
\end{equation}
In SI, each speaker is represented by the a GMM and is referred to by his/her model $\lambda$. The parameter of $\lambda$ are optimized using expectation maximization(EM) algorithm \cite{demp}. In these experiments, the GMMs are trained with 10 iterations where clusters are initialized by vector quantization \cite{lbg} algorithm.

In identification stage, the log-likelihood scores of the feature vector of the utterance under test is calculated by
\begin{equation}\label{gmm4}
\log p(\boldsymbol{X}|\lambda)=\sum_{t=1}^{T}p(\boldsymbol{x}_t|\lambda)
\end{equation}
Where $\textbf{X}=\{\boldsymbol{x}_1,\boldsymbol{x}_2,...,\boldsymbol{x}_t \}$ is the feature vector of the test utterance.

In closed set SI task, an unknown utterance is identified as an utterance of a particular speaker whose model gives maximum log-likelihood.
It can be written as
\begin{equation}\label{gmm5}
\hat{S}=\arg \mathop {\max }\limits_{1 \le k \le S}\sum_{t=1}^{T}p(\boldsymbol{x}_t|\lambda_{k})
\end{equation}
where $\hat{S}$ is the identified speaker from speaker's model set $\Lambda =\{\lambda_1,\lambda_2,...,\lambda_S\}$ and $S$ is the total number of speakers.

\subsubsection{Databases for experiments}

\hspace{1cm}

\emph{YOHO Database:} The YOHO voice verification corpus \cite{camp,yoho} was collected while
testing ITT's prototype speaker verification system in an office
environment. Most subjects were from the New York City area,
although there were many exceptions, including some non-native
English speakers. A high-quality telephone handset (Shure XTH-383)
was used to collect the speech; however, the speech was not passed
through a telephone channel. There are $138$ speakers ($106$ males
and $32$ females); for each speaker, there are $4$ enrollment
sessions of $24$ utterances each and $10$ test sessions of $4$
utterances each. In this work, a closed set text-independent
speaker identification problem is attempted where we consider all
$138$ speakers as client speakers. For a speaker, all the $96\ (4\
\mathrm{sessions}\times24\ \mathrm{utterances})$ utterances are
used for developing the speaker model while for testing, $40\ (10\
\mathrm{sessions}\times 4 \ \mathrm{utterances})$ utterances are
put under test. Therefore, for $138$ speakers we put
$138\times40=5520$ utterances under test and evaluated the
identification accuracies.
\par
\emph{POLYCOST Database:} The POLYCOST database \cite{melin} was recorded as a common
initiative within the COST $250$ action during January- March
1996. It contains around $10$ sessions recorded by $134$ subjects
from $14$ countries. Each session consists of $14$ items, two of
which (MOT01 \& MOT02 files) contain speech in the subject's
mother tongue. The database was collected through the European
telephone network. The recording has been performed with ISDN
cards on two XTL SUN platforms with an $8$ kHz sampling rate. In
this work, a closed set text independent speaker identification
problem is addressed where only the mother tongue (MOT) files are
used. Specified guideline \cite{melin} for conducting closed set
speaker identification experiments is adhered to, i.e. `MOT02'
files from first four sessions are used to build a speaker model
while `MOT01' files from session five onwards are taken for
testing. As with YOHO database, all speakers ($131$ after deletion
of three speakers) in the database were registered as clients.

\subsubsection{Score Calculation}
In closed-set speaker identification problem, identification
accuracy as defined in \cite{rey} and given by the equation (\ref{eqpia}) is followed.
\begin{eqnarray}\label{eqpia}
\mathrm{Percentage\ of\ identification\ accuracy\
(\textbf{\textbf{PIA}})}&=&\nonumber\\
\frac{\mathrm{No.\ of\ utterance\ correctly\
identified}}{\mathrm{Total\ no.\ of\ utterance\ under\
test}}\times 100\label{poc}
\end{eqnarray}

\subsection{Speaker Identification Experiments and Results}
Experiments were performed using GMM based classifier of different model orders which are power of two i.e. $2$, $4$, $8$, $16$, etc. The number of gaussian is limited by the amount of available training data (Average training speech length per speaker after silence removal: (i) POLYCOST-40s and (ii) YOHO-150s). The number of mixtures are incremented to $16$ for POLYCOST and $64$ for YOHO database.
We have evaluated the performance of ACRLAG feature as a front end for speaker identification task for both the databases. Experiments were conducted for different order of linear prediction and lag for optimal performance. An LP order of 12-20 is sufficient in capturing speaker specific information from the residual signal. Exhaustive search was also carried out for finding the optimal value of lag. The lag was chosen so that it can effectively capture the second order properties (autocorrelation) of one pitch (or pitch like information) in the residual signal. Experimentally we have observed that autocorrelation computation with a lag of 10-16 is sufficient to represent the residual information.
In Table \ref{tableacflag} the SI performance using ACRLAG is shown for LP order of 13 and 12 lags. Hence the number of residual feature becomes 12+1=13. The performance attained through only residual feature (vocal cord information) is not much considerable, but it contains some useful complementary information which cannot be explained by standard spectral features. Hence, the LLR of the two systems are combined as stated in \ref{fusion}. In our experiments the dimension for various spectral features is set to 19 for fare comparison, and this order is frequently used to model spectral information.  The LP based features (LPCC, LAR, LSF, PLPCC) are extracted using 19 order all pole modeling; on the other hand the filterbank based features (LFCC and MFCC) are extracted using 20 triangular shaped bandpass filters which are equally spaced either in Hertz scale (for LFCC) or in Mel scale (for MFCC). In Table \ref{tablepoly} and \ref{tableyoho} the PIA for baseline systems as well as the fused systems (with $\eta=0.5$) are shown. The performance of the fused system is better throughout different spectral features and different model orders of GMM for both the databases. The improvement in performance is higher in lower model order compared to higher model order. This is due to base effect which is usually experienced in performance analysis of newly proposed features for a classifier system. For example, in case of PLPCC feature based SI system performance is improved by $12\%$ for POLYCOST and $15.13\%$ for YOHO database using model order 2. But, in case of higher model order the improvements are $4.74\%$ and $1.47\%$. The improvement in POLYCOST database (telephonic) is also significantly better than that of YOHO (microphonic) for various features. In this work all voiced and unvoiced speech frames are utilized for extracting residual features. Though only voiced frames contain pitch and significant residual information, still unvoiced frames contain information which is not considered in auto regressive (AR) based approaches (LP). Filter bank based approaches also removes pitch or pitch like information when wrapping with triangular filters.  We have also performed an experiment by picking only voiced frames and observed that taking  only those frames the performance is not improved, rather degraded due to the less amount of training data and curse of dimensionality effect in GMM modeling.
\par
It is desirable to note that the performance of the combined system which essentially considers 19+13=32 dimensions is better than the system which is based on any 32 dimensional single spectral feature based system.

\begin{table}[htbp]
\begin{center}
\caption{Speaker Identification Results on POLYCOST and YOHO database using ACRLAG feature for different model order of GMM (ACRLAG Configuration: LP Order $=13$, Number of Lag$=12$).}
\begin{tabular}{|c|c|c|}
\hline
\hline
Database &
No. of Mixtures&
Identification Accuracy \\
\hline
\hline
\raisebox{-4.50ex}[0cm][0cm]{POLYCOST}&
2&  45.7560\\
\cline{2-3}
 & 4& 49.3369 \\
\cline{2-3}
 & 8& 55.7029 \\
\cline{2-3}
 & 16&  59.6817 \\
\hline
\hline
\raisebox{-4.50ex}[0cm][0cm]{YOHO}&
2&  41.2319\\
\cline{2-3}
 & 4&  49.3116\\
\cline{2-3}
 & 8&   56.9022\\
\cline{2-3}
 & 16&  63.2246\\
\cline{2-3}
& 32&  68.9130\\
\cline{2-3}
 & 64&  73.3514\\
\hline
\hline
\end{tabular}
\label{tableacflag}
\end{center}
\end{table}

\begin{table}[htbp]
\begin{center}
\caption{Speaker Identification Results on POLYCOST database showing the performance of baseline (single stream) system and fused system.)}
\begin{tabular}{|c|c|c|c|c|}
\hline
\hline
Spectral & No. of & Baseline& Fused \\
Feature  & Mixtures & System & System\\
\hline
\hline
\raisebox{-4.50ex}[0cm][0cm]{LPCC}&
2& 63.5279& \textbf{71.2202} \\
\cline{2-4}
 & 4& 74.5358& \textbf{79.8408} \\
\cline{2-4}
 & 8& 80.3714& \textbf{83.2891} \\
\cline{2-4}
 & 16& 79.8408& \textbf{83.0239} \\
\hline
\hline
\raisebox{-4.50ex}[0cm][0cm]{LAR}&
2& 62.3342& \textbf{66.8435} \\
\cline{2-4}
 & 4& 72.5464& \textbf{75.8621} \\
\cline{2-4}
 & 8& 76.2599& \textbf{79.0451} \\
\cline{2-4}
 & 16& 78.6472& \textbf{80.2387} \\
\hline
\hline
\raisebox{-4.50ex}[0cm][0cm]{LSF}&
2& 60.7427& \textbf{67.7719} \\
\cline{2-4}
 & 4& 66.8435& \textbf{72.1485}\\
\cline{2-4}
 & 8& 75.7294& \textbf{78.9125} \\
\cline{2-4}
 & 16& 78.1167& \textbf{80.9019} \\
\hline
\hline
\raisebox{-4.50ex}[0cm][0cm]{PLPCC}&
2& 62.9973& \textbf{70.5570} \\
\cline{2-4}
 & 4& 72.2812& \textbf{77.0557} \\
\cline{2-4}
 & 8& 75.0663& \textbf{80.1061} \\
\cline{2-4}
 & 16& 78.3820& \textbf{82.0955} \\
\hline
\hline
\raisebox{-4.50ex}[0cm][0cm]{LFCC}&
2& 62.7321& \textbf{72.1485} \\
\cline{2-4}
 & 4& 74.9337& \textbf{77.7188} \\
\cline{2-4}
 & 8& 79.0451& \textbf{82.3607} \\
\cline{2-4}
 & 16& 80.7692& \textbf{84.2175} \\
\hline
\hline
\raisebox{-4.50ex}[0cm][0cm]{MFCC}&
2& 63.9257&\textbf{70.1592} \\
\cline{2-4}
 & 4& 72.9443& \textbf{75.7294} \\
\cline{2-4}
 & 8& 77.8515& \textbf{79.3103} \\
\cline{2-4}
 & 16& 77.8515& \textbf{79.8408} \\
\hline
\hline
\end{tabular}
\label{tablepoly}
\end{center}
\end{table}

\begin{table}[htbp]
\begin{center}
\caption{Speaker Identification Results on YOHO database showing the performance of baseline (single stream) system and fused system.)}
\begin{tabular}{|c|c|c|c|}
\hline
\hline
Spectral & No. of & Baseline& Fused \\
Feature  & Mixtures & System & System\\
\hline
\hline
\raisebox{-7.50ex}[0cm][0cm]{LPCC}&
2& 80.9964& \textbf{82.8986} \\
\cline{2-4}
& 4& 89.0036& \textbf{90.3261} \\
\cline{2-4}
 & 8& 93.9312& \textbf{94.7101} \\
\cline{2-4}
 & 16& 95.7246& \textbf{96.0507} \\
\cline{2-4}
 & 32& 96.4855& \textbf{96.9746} \\
\cline{2-4}
 & 64& 96.8297& \textbf{97.3007} \\
\hline
\hline
\raisebox{-7.50ex}[0cm][0cm]{LAR}&
2& 72.6449& \textbf{74.2935}\\
\cline{2-4}
& 4& 80.9601& \textbf{82.3370} \\
\cline{2-4}
 & 8& 87.4094& \textbf{88.8587} \\
\cline{2-4}
 & 16& 91.4855& \textbf{92.4094} \\
\cline{2-4}
 & 32&  93.2246& \textbf{93.8949} \\
\cline{2-4}
 & 64& 94.7826& \textbf{94.9819} \\
\hline
\hline
\raisebox{-7.50ex}[0cm][0cm]{LSF}&
2& 70.7428& \textbf{77.5181} \\
\cline{2-4}
& 4& 81.3768& \textbf{85.2899} \\
\cline{2-4}
 & 8& 90.4529& \textbf{91.9203} \\
\cline{2-4}
 & 16& 93.2246& \textbf{94.3841} \\
\cline{2-4}
 & 32& 95.5978& \textbf{95.9964} \\
\cline{2-4}
 & 64& 96.5761& \textbf{97.0290} \\
\hline
\hline
\raisebox{-7.50ex}[0cm][0cm]{PLPCC}&
2& 66.5761& \textbf{76.6486} \\
\cline{2-4}
 & 4& 76.9203& \textbf{83.8225} \\
\cline{2-4}
 & 8& 85.3080& \textbf{90.1449} \\
\cline{2-4}
 & 16& 90.6341& \textbf{93.1522} \\
\cline{2-4}
 & 32& 93.5326& \textbf{95.3804} \\
\cline{2-4}
 & 64& 94.6920& \textbf{96.0870} \\
\hline
\hline
\raisebox{-7.50ex}[0cm][0cm]{LFCC}&
2& 83.0072& \textbf{83.2609} \\
\cline{2-4}
 & 4& 90.3623&  \textbf{90.6703} \\
\cline{2-4}
 & 8& 94.6196& \textbf{ 95.2536} \\
\cline{2-4}
 & 16& 96.2681& \textbf{96.5942} \\
\cline{2-4}
 & 32& 97.1014& \textbf{97.1377} \\
\cline{2-4}
 & 64& 97.2464& \textbf{97.3732} \\
\hline
\hline
\raisebox{-7.50ex}[0cm][0cm]{MFCC}&
2& 74.3116& \textbf{80.4891} \\
\cline{2-4}
 & 4& 84.8551& \textbf{87.8080} \\
\cline{2-4}
 & 8& 90.6703& \textbf{92.9167} \\
\cline{2-4}
 & 16& 94.1667& \textbf{95.1993} \\
\cline{2-4}
 & 32& 95.6522& \textbf{96.3587} \\
\cline{2-4}
 & 64& 96.7935& \textbf{97.2101} \\
\hline
\hline
\end{tabular}
\label{tableyoho}
\end{center}
\end{table}

\section{Conclusion}
In this paper we present a new scheme for representing vocal cord characteristics. The vocal source feature is extracted using autocorrelation of the residual signal with some lag. The complementarity of this proposed feature with short term spectral feature is exploited through fusion technique. The linear combination of log likelihood ratio score is formulated to utilize the advantages for combined system. The performance can further be enhanced with the help of advanced fusion techniques and optimal selection of LP order and lag.

\bibliography{latexbib}

\end{document}